\documentstyle[preprint,aps,floats,epsf]{revtex}

\begin{document}

\title{Narrow Pentaquark States in a Quark Model with 
Antisymmetrized Molecular Dynamics
}

\author{Y. Kanada-En'yo, O. Morimatsu and T. Nishikawa}

\address{Institute of Particle and Nuclear Studies,\\
High Energy Accelerator Research Organization,\\
1-1 Oho, Tsukuba, Ibaraki 305-0801, Japan}

\maketitle

\begin{abstract}
The exotic baryon $\Theta^+(uudd\bar{s})$ is studied with microscopic 
calculations in a quark model 
by using a method of antisymmetrized molecular dynamics(AMD).
We predict narrow states, $J^\pi=1/2^+(I=0)$,
$J^\pi=3/2^+(I=0)$, and $J^\pi=3/2^-(I=1)$, which 
nearly degenerate in a low-energy region of the $uudd\bar{s}$ system.
We discuss $NK$ decay widths and estimate them to be $\Gamma< 7$ for 
the $J^\pi=\{1/2^+,3/2^+\}$, and $\Gamma<1$ MeV for the $J^\pi=3/2^-$ state.
\end{abstract}

The evidence of an exotic baryon $\Theta^+$ has recently been reported 
by several experimental 
groups.
This discovery proved the existence 
of the multiquark hadron, whose minimal quark content is $uudd\bar{s}$
as given by the decay modes.
The study of pentaquarks has become a hot subject in hadron physics.
A chiral soliton model \cite{diakonov} predicted a narrow 
$\Theta^+$($J^\pi=1/2^+$) state 
whose parity contradicts the naive quark model expectation.
Theoretical studies were done to describe $\Theta^+$ 
by many groups\cite{jaffe,oka04}.
The spin parity of $\Theta^+$ is not only a open problem but
also a key property to understand the dynamics of pentaquark systems. 

In this paper we would like to clarify the mechanism 
of the existence of narrow pentaquark states.
We try to extract a simple picture for the pentaquark 
baryon with levels, width, spin-parity and structure
from explicit calculation.
In order to achieve this goal, we study the pentaquark 
with a flux-tube model\cite{carlson,morimatsu} based on strong coupling QCD,
by using a AMD method\cite{ENYObc,AMDrev}.

\begin{figure}[h]
\noindent
\epsfxsize=0.9\textwidth
\centerline{\epsffile{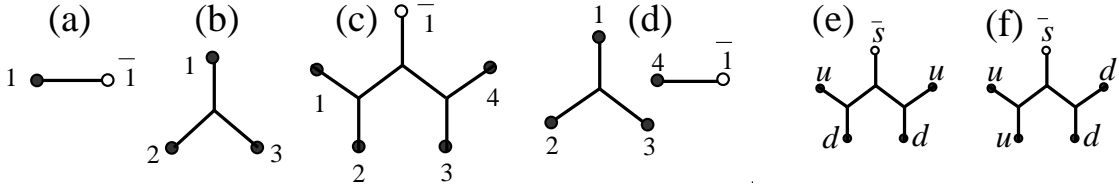}}
\caption{\label{fig:flux}
Flux-tube configurations for confined states of $q\bar{q}$ (a),
$q^3$ (b), $q^4\bar{q}$ (c), and disconnected flux-tube of
$q^4\bar{q}$ (d). Figures (e) and (f) represent the flux tubes
in the color configurations, 
$[ud][ud]\bar{s}$ and
$[uu][dd]\bar{s}$, respectively.
The string potentials given by the flux tubes (b) and (c)
are supported by Lattice QCD \protect\cite{Takahashi}.}
\end{figure}

In the flux-tube model, the interaction energy of quarks 
and anti-quarks is given by the energy of the string-like 
color-electric flux, which is proportional to the minimal 
length of the flux-tube connecting quarks and anti-quarks 
at long distances supplemented by perturbative one-gluon-exchange (OGE) 
interaction at short distances.
For the $q^4\bar q$ system the flux-tube configuration 
has an exotic topology, Fig.\ref{fig:flux}(c), 
in addition to an ordinary meson-baryon
topology, Fig. \ref{fig:flux}(d).
An important point is that 
the transition between the different flux-tube 
topologies (c) and (d) is strongly suppressed because it takes place 
only in higher order.
(In 1991, Carlson and Pandharipande studied exotic 
hadrons in the flux-tube model\cite{carlsonb} 
and calculated a few $q^4\bar q$ states with very limited quantum numbers.)

We apply the AMD method to the flux-tube model and calculate
the $uudd\bar{s}$ system.
The AMD is a variational method to solve a finite many-fermion system.
One of the advantages of this method is that 
the spatial and spin degrees of freedom  
for all particles are independently treated.
This method can successfully describe various types of structure such as 
shell-model-like structure and clustering (correlated nucleons)
in nuclear physics\cite{ENYObc,AMDrev}. 
With the AMD method we calculate all the possible spin parity states 
of $uudd\bar{s}$
system, and analyze the wave function to 
estimate the decay widths of the obtained states 
with a method of reduced width amplitudes.

In the present calculation, the quarks are treated as 
non-relativistic spin-$\frac{1}{2}$ Fermions.
We use a Hamiltonian as $H=H_0+H_I+H_f$,
where $H_0$ is the kinetic energy of 
the quarks, $H_I$
represents the short-range OGE interaction between the quarks
and $H_f$ is the energy of the flux tubes.
$H_0$ and $H_I$ are represented as follows;
\begin{eqnarray}
H_0 & =& \sum_{i}m_i+\sum_{i}\frac{{p}_i^2}{2m_q}-T_0,\\
H_I & =& \alpha_c\sum_{i<j}F^\alpha_i F^\alpha_j
\left[\frac{1}{r_{ij}}-\frac{2\pi}{3m_i m_j}s(r_{ij})
\sigma_i\cdot\sigma_j\right],
\end{eqnarray}
where $m_i$(the $i$-th quark mass) is $m_q$ for a $u$ or $d$ quark
and $m_s$ for a $\bar{s}$ quark, and
$T_0$ denotes the kinetic energy of the center-of-mass motion.
Here, we do not take into account the mass difference 
between $ud$ and $s$ 
in the second term of $H_0$, for simplicity. 
$\alpha_c$ is the quark-gluon coupling constant, and 
$F^\alpha_i$ is the 
generator of color $SU(3)$.
In $H_I$, we take only the dominant terms, Coulomb and color-magnetic terms,
and omit other terms.

In the flux-tube quark model \cite{carlson},
the confining potential is written as
$H_{f}={\sigma}L_f-M^0$,
where ${\sigma}$ is the string tension, $L_f$ is
the minimum length of the flux tubes, and $M^0$ is the 
zero-point energy. 
$M^0$ depends on the topology of the flux tubes 
and is necessary to fit the $q\bar{q}$, $q^3$ and $q^4\bar{q}$ 
potential. In the present calculation, we adjust the $M^0$ to 
fit the absolute masses for each of $3q$-baryon and pentaquark.
For the meson and 3$q$-baryon systems, the flux tube configurations 
are given as Fig.\ref{fig:flux}(a) and (b).
For the pentaquark system, the different types of 
flux-tube configurations 
appear as shown in Fig.\ref{fig:flux}(e),(f), and (d), which correspond
to the states,
$|\Phi_{(e)}\rangle=|[ud][ud]\bar{s}\rangle$,
$|\Phi_{(f)}\rangle=|[uu][dd]\bar{s}\rangle$,
and $|\Phi_{(d)}\rangle=|(qqq)_{1}(qq)_1\rangle$,
respectively ($[qq]$ is defined by color anti-triplet of $qq$). 
In the present calculation of energy variation, 
we neglect the transitions among $|\Phi_{(e)}\rangle$, 
$|\Phi_{(f)}\rangle$ and $|\Phi_{(d)}\rangle$ and solve 5$q$ wave functions
within the model space (e) or (f), which corresponds to the confined states.
It is reasonable because the transitions are suppressed as mentioned before.
In the practical calculations of the string potential
$\langle\Phi|H_f|\Phi\rangle$, 
the minimum length of the flux tubes $L_f$ is approximated by
a linear combination of two-body distances as 
$L_f\approx \frac{1}{2}(r_{12}+r_{23}+r_{31})$ for a $3q$-baryon,
 and $L_f\approx \frac{1}{2}(r_{12}+r_{34})+
\frac{1}{8}(r_{13}+r_{14}+r_{23}+r_{24})
+\frac{1}{4}(r_{\bar{1}1}+r_{\bar{1}2}+r_{\bar{1}3}+r_{\bar{1}4})$
for $\Phi_{(e)}$ or $\Phi_{(f)}$ of the pentaquark systems.
We note that the confinement is reasonably realized by the 
approximation for $\Phi_{(e,f)}$ as follows.
The flux-tube configuration (e)(or (f)) consists of seven bonds and 
three junctions. In the limit that the length($R$) of any one bond becomes 
much larger than other bonds, the approximated $\langle H_f \rangle$ 
behaves as a linear potential $\sigma R$.
It means that all the quarks and anti-quarks are 
bounded by the linear potential with the tension $\sigma$.
Therefore, the approximation for $\Phi_{(e)}$ or $\Phi_{(f)}$
is a natural extension of the usual approximation for $3q$-baryons. 
It is easily proved that the approximations 
are equivalent to 
$\langle\Phi|H_f|\Phi\rangle\approx \langle\Phi|{\hat O}|\Phi\rangle$,
where ${\hat O}\equiv -\frac{3}{4}\sigma\sum_{i<j}F^\alpha_i 
F^\alpha_j r_{ij}-M^0$, 
within each of the flux-tube configurations.

We solve the eigenstates of the Hamiltonian with a variational method
in the AMD model space\cite{ENYObc,AMDrev}. 
We take a base AMD wave function in a quark model 
as follows. 
\begin{equation}
\Phi({\bf Z})=(1\pm P) {A}
\left[\phi_{Z_1}\phi_{Z_2}\cdots\phi_{Z_{N_q}} X \right],
\ \ \ \  \phi_{Z_i}\propto
e^{-\frac{1}{2b^2}(r-\sqrt{2}bZ_i)^2},
\end{equation}
where $1\pm P$ is the parity projection operator, $A$ is the
anti-symmetrization operator,  
and the spatial part $\phi_{Z_i}$ of 
the $i$-th single-particle wave function is
written by a Gaussian with the center $Z_i$($Z_i$ is 
a complex parameter). $X$ is the spin-isospin-color function.
For the pentaquark($uudd\bar{s}$) system, $X$ is expressed as
\begin{eqnarray}
X=& \sum_{\footnotesize m_{1},m_{2},m_{3},m_{4},m_{5}} 
c_{m_{1}m_{2}m_{3}m_{4}m_{5}}
|m_{1}m_{2}m_{3}m_{4}m_{5}\rangle_S \nonumber\\
& \otimes \{ |udud\bar{s}\rangle {\rm \ or\ }
 |uudd\bar{s}\rangle \}\otimes  
\epsilon_{abg}\epsilon_{ceh}\epsilon_{ghf}|abce\bar{f}\rangle_C,
\end{eqnarray}
where $|udud\bar{s}\rangle$ and $|uudd\bar{s}\rangle$ 
correspond to the configurations $[ud][ud]\bar{s}$ and 
$[uu][dd]\bar{s}$ in Fig.\ref{fig:flux}, respectively.
Here, $|a\rangle_C (a=1,2,3)$ denotes the color function,
and $|m\rangle_S (m=\uparrow,\downarrow)$ 
is the intrinsic-spin function. 
Since we are interested in the confined states, 
we adopt those model space for 
the color configurations $(qq)_{\bar{3}}(qq)_{\bar{3}}\bar{q}$,
but do not use the meson-baryon configurations $(qqq)_{1}(q\bar{q})_{1}$.
The variational parameters are ${\bf Z}=\{Z_1,Z_2,\cdots,Z_5\}$ and
$c_{m_{1}m_{2}m_{3}m_{4}m_{5}}$ which specify the 
spatial and spin configurations. The energy variation for ${\bf Z}$
is performed by  a frictional cooling method,
and the coefficients $c_{m_{1}m_{2}m_{3}m_{4}m_{5}}$ 
are determined by diagonalization of Hamiltonian and norm
matrices. After the energy variation, the intrinsic-spin and parity $S^\pi$ 
eigen wave function $\Phi({\bf Z})$ for the lowest state 
is obtained for each $S^\pi$. 

In the numerical calculation, the linear and Coulomb potentials
are approximated by seven-range Gaussians.
We use the parameters,
$\alpha_c=1.05$,
$\Lambda=0.13$ fm,
$m_q=0.313$ GeV,
${\sigma}=0.853$ GeV/fm, and
$\Delta m_s=m_s-m_q=0.2$ GeV.
The quark-gluon coupling constant $\alpha_c$ is 
chosen so as to fit the $N$ and $\Delta$ 
mass difference.
The string tension $\sigma$ is
adopted to adjust the excitation energy of $N^*(1520)$.
The width parameter $b$ is chosen to be $0.5$ fm.
By choosing $M_0$ as $M^0_{q^3}=972$ MeV, the masses of
$N$, $N^*(1520)$ and $\Delta$
are fitted \cite{ENYO-penta}, and the masses of $\Lambda$,
$\Sigma$ and $\Sigma^*{1385}$ are well reproduced with these parameters.

Now, we apply the AMD method to the $uudd\bar{s}$ system.
For each spin parity, we calculate energies of the  
 $[ud][ud]\bar{s}$ and 
$[uu][dd]\bar{s}$ states
and adopt the lower one.
In table.\ref{tab:5q}, the calculated results are shown.
We adjust the zero-point energy of the string potential $M_0$
as $M^0_{q^4\bar{q}}=2385$ MeV 
to fit the absolute mass of the recently observed $\Theta^+$.
This $M^0_{q^4\bar{q}}$ for pentaquark system is chosen
independently of $M^0_{q^3}$ for $3q$-baryon.
If $M^0_{q^4\bar{q}}=\frac{5}{3}M^0_{q^3}$ is assumed 
as Ref.[8],
the calculated mass of the pentaquark
is around 2.2 GeV, which is consistent with the result of 
Ref.[8].

The most striking point in the results is that 
the $S^\pi=3/2^-$ and $S^\pi=1/2^+$ states nearly degenerate
with the $S^\pi=1/2^-$ states.
The $S^\pi=1/2^+$ correspond to $J^\pi=1/2^+$ and $3/2^+$ 
with $S=1/2,L=1$, and the $S^\pi=3/2^-$ is 
$J^\pi=3/2^-$($S=3/2,L=0$).
The lowest state $J^\pi=1/2^-(S^\pi=1/2^-,L=0$) exists just below the 
$J^\pi=3/2^-$ state, however, 
this state, as we discuss later, is expected to be 
much broader than other states. 
Other spin-parity states are much higher than these
low-lying states.

The $LS$-partners, $J^\pi=1/2^+$ and $3/2^+$ exactly degenerate
in the present Hamiltonian where the spin-orbit and tensor terms are omitted.
If we introduce the spin-orbit force into the Hamiltonian
the $LS$-splitting is small in the diquark structure
because the effect of the spin-orbit force from the 
spin-zero diquarks is very weak as discussed in Ref.\cite{Close}. 
As shown later, 
since the present results show that the diquark structure is realized in
the $J^\pi=1/2^+$ and $3/2^+$ states, the $LS$-splitting should not be large
in the $uudd\bar{s}$ system.

\begin{table}[ht]
\begin{tabular}{@{}rrrrrr@{}}
& $[uu][dd]\bar{s}$
& $[ud][ud]\bar{s}$
& $[ud][ud]\bar{s}$
& $[ud][ud]\bar{s}$
& $[uu][dd]\bar{s}$\\
$S^\pi$ & $\frac{1}{2}^-$& $\frac{3}{2}^-$& $\frac{1}{2}^+$& $\frac{1}{2}^-$&
 $\frac{5}{2}^-$\\
\hline
Kinetic($H_0$) &3.23 &3.22 	&3.36 	&3.19 	&3.19 	\\
String($H_F$) & $-0.67$ &$-$0.66 	&$-$0.55&$-$0.64&$-$0.64 \\
Coulomb&$-$1.05&$-$1.04&$-$0.99&$-$1.03&$-$1.03\\
Color mag.&$-0.01$ &0.01 	&$-$0.25&0.04 	&0.19 	\\
$q\bar{q}$Color mag. &$-0.06$ &$-$0.01&0.00 	&0.02 	&0.06 	\\
\hline
$E$ & 1.50 &1.53 	&1.56 	&1.56 	&1.71 	\\
\end{tabular}
\caption{Calculated masses(GeV) of the $uudd\bar{s}$ system.
The expectation values of the 
kinetic, string, Coulomb, color-magnetic terms, and that 
of the color-magnetic term in $q\bar{q}$ pairs are listed.
The $S^\pi=3/2^+$ and $S^\pi=5/2^+$ states are 
higher than the $S^\pi=5/2^-$ state.\label{tab:5q}
}
\end{table}

Next, we analyze the spin structure of these states, and found that
the $J^\pi=\{1/2^+,3/2^+\}(S=1/2,L=1)$ states consist of 
two spin-zero $ud$-diquarks, while 
the $J^\pi=3/2^-$ consists
 of a spin-zero $ud$-diquark and a spin-one
$ud$-diquark. 
Since the spin-zero $ud$-diquark has the isospin $I=0$ and 
the spin-one $ud$-diquark has $I=1$ 
because of the color asymmetry, the isospin of the 
$J^\pi=3/2^-$ state is $I=1$, while 
the even-parity states $J^\pi=1/2^+,3/2^+$ are
$I=0$. We consider that the $J^\pi=1/2^+$ state corresponds 
to the $\Theta^+$(1530) in the flavor ${\overline{10}}$-plet 
predicted by Diakonov et al.\cite{diakonov}.
The odd-parity state, $J^\pi=3/2^-$ 
has $I=1$, which means that this state is a 
member of the flavor ${27}$-plet.
We denote the $J^\pi=1/2^+,3/2^+(I=0)$ 
by $\Theta^+_0$, and the $J^\pi=3/2^-(I=1)$ 
by $\Theta^+_1$.

Although it is naively expected that unnatural spin parity states 
are much higher than the natural spin-parity $1/2^-$ state, 
the results show the abnormal 
level structure of the $(udud\bar{s})$ system, where 
the high spin state, $J^\pi=3/2^-$, and the 
unnatural parity states, $J^\pi=\{1/2^+,3/2^+\}$, nearly degenerate
just above the $J^\pi=1/2^-$ state. 
By analysing the details of these states,
the abnormal level structure can be easily understood with a simple picture
as follows.
As shown in table.\ref{tab:5q}, 
the $J^\pi=\{1/2^+,3/2^+\}(S=1/2,L=1)$ states have larger
kinetic and string energies
than the $J^\pi=3/2^-(S=3/2,L=0)$ and $J^\pi=1/2^-(S=1/2,L=0)$
states, while the former states
gain the color-magnetic interaction. It indicates that the degeneracy of 
the even-parity states with the odd-parity states
is realized by the balance of the loss of kinetic and 
string energies and the gain of the color-magnetic interaction.
In the $J^\pi=\{1/2^+,3/2^+\}$ and the $3/2^-$ states, 
the competition of the energy loss and gain 
can be simply understood 
from the point of view of the diquark structure as follows. 
As already mentioned by Jaffe and Wilczek\cite{jaffe}, the relative motion 
between two spin-zero diquarks must have the odd parity ($L=1$)
because of Pauli blocking between the two identical diquarks. 
In the $J^\pi=3/2^-$ state, 
one of the spin-zero $ud$-diquarks is broken to be a 
spin-one $ud$-diquark to avoid the Pauli blocking, then,
the $L=0$ is allowed because diquarks are not identical. 
The $L=0$ is energetically favored in the kinetic and string terms, and 
the energy gain cancels the color-magnetic energy loss of a 
spin-one $ud$-diquark. 
Although we can not describe the $J^\pi=1/2^-$ state by 
such a simple diquark picture, the competition of energy loss and gain 
in this state is similar to the $J^\pi=3/2^-$.

\begin{figure}
\noindent
\epsfxsize=0.75\textwidth
\centerline{\epsffile{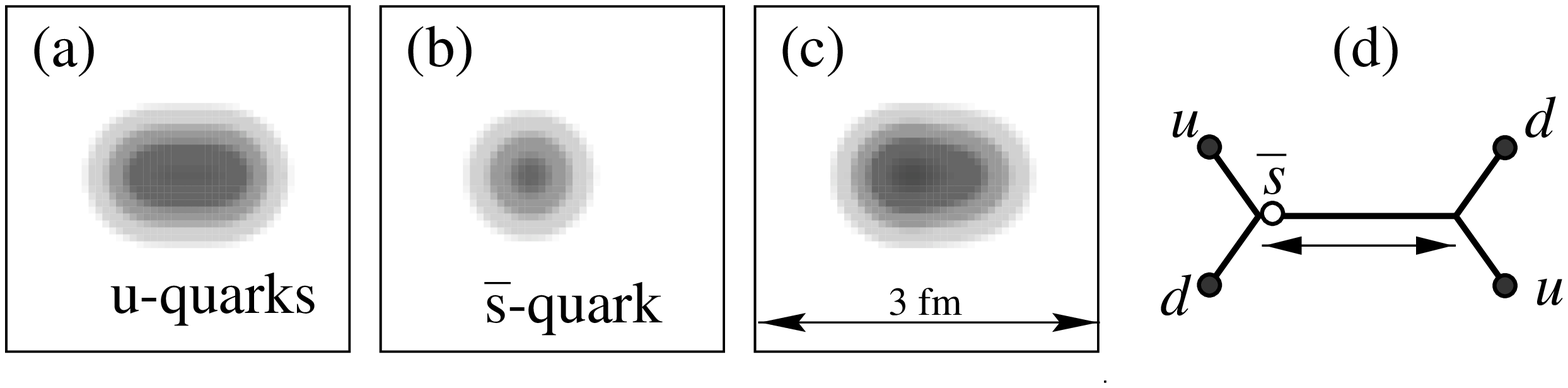}}
\caption{\label{fig:5q}
$q$ and $\bar{q}$ density distribution in the 
$J^\pi=1/2^+,3/2^+(S=1/2,L=1)$ states of $Theta^+$.
The $u$-quark density (a), $\bar{s}$ density (b), and 
total quark-antiquark density
(c) of the intrinsic state before parity projection are shown. 
The $d$-quark density is same as the $u$-quark density.
 The root-mean-square
radius of $q$ and $\bar{q}$ is 0.63 fm (the nucleon size is 0.5 fm). 
}
\end{figure}

We remark that the existence of two spin-zero $ud$-diquarks in
the $J^\pi=\{1/2^+,3/2^+\}$ states predicted 
by Jaffe and Wilczek\cite{jaffe} is actually confirmed 
in our {\it ab initio} calculations.
We found that the component 
with two spin-zero $ud$-diquarks is 97\% in the present 
$J^\pi=\{1/2^+,3/2^+\}$ state.
In Fig.\ref{fig:5q}, we show the quark and 
anti-quark density distributions in the
$J^\pi=\{1/2^+,3/2^+\}$ states.
In the intrinsic state before parity projection, 
we found the spatial development of $ud$-$uds$ clustering, which causes
a parity-asymmetric shape (Fig.\ref{fig:5q} (c)).

We estimate the $KN$-decay widths of these states by using a method 
of reduced width amplitudes\cite{Horiuchi73}. 
The decay width $\Gamma$
is estimated by the product $\Gamma^0_L\times S_{fac}$, where
$\Gamma^0_L(a,E_{th})$ is given by the penetrability of the 
barrier\cite{ENYO-penta},
and $S_{fac}(a)$ is the probability of the decaying particle 
at the channel radius $a$. 
In the following discussion,
we use the channel radius $a=1$ fm and the threshold energy $E_{th}=100$ MeV.
We here estimate the maximum values of the widths, by taking into account
only quark degrees of freedom.
We omit the suppression of the transition 
between the confined state and the meson-baryon state due to
the rearrangement of flux-tubes, which makes  
$S_{fac}$ small in general.

In case of even parity $J^\pi=1/2^+, 3/2^+$ states, the $KN$ decay modes 
are the $P$-wave, which gives $\Gamma^0_{L=1}\approx 100$ MeV\ fm$^{-1}$.
We calculate the overlap between the obtained pentaquark
wave function and the $K^+n$ state, and evaluate the probability 
as $S_{fac}=0.034$ fm$^{-1}$.
Roughly speaking, the main factors in this meson-baryon probability 
are the factor $\frac{1}{3}$ from the color configuration, 
the factor $\frac{1}{4}$ from the intrinsic spin part, and
the other factor which arises from the spatial overlap.
By using this value, the total width for 
$K^+n$ and $K^0p$ decays of the $J^\pi=1/2^+,3/2^+$ states 
is estimated to be $\Gamma < 7$ MeV.
For more quantitative discussions, 
it is important to treat the coupling with the $KN$ continuum states,
where one must take into account 
the suppression due to the rearrangement of 
flux-tube topologies.

It is interesting that the $KN$ decay width of 
the $J^\pi=3/2^-$ state is extremely small due to the 
$D$-wave centrifugal barrier. In fact,
$\Gamma^0_{L=2}\approx 30$ MeV\ fm$^{-1}$ is much smaller 
than the $P$-wave case.
Moreover, the $J^\pi=3/2^-$($S^\pi=3/2^-$,$L=0$) 
has no $D$-wave component, therefore, no overlap with the $KN$($L=2$) states
in the present calculation. 
Even if we introduce the spin-orbit or tensor forces,
the $KN$ probability($S_{fac}$) in the $J^\pi=3/2^-$ pentaquark state
is expected to be minor.
Consequently, the $J^\pi=3/2^-$ state should be extremely narrow.
If we assume the $S_{fac}$ in the $J^\pi=3/2^-$ to be half of that in
the $J^\pi=1/2^+, 3/2^+$ states,
the $KN$ decay width is estimated to be $\Gamma<1$ MeV.
Contrary to the narrow features of the $J^\pi=3/2^-$ state, 
in case of $J^\pi=1/2^-$, $S$-wave($L=0$) decay is allowed 
and this state should be much broader.

In conclusion, we proposed a quark model in the framework of the AMD method,
and applied it to the $uudd\bar{s}$ system. 
The level structure of the the $uudd\bar{s}$ system 
and the properties of the low-lying states were studied. 
We predicted that the narrow 
$J^\pi=\{1/2^+$,$3/2^+$\}($\Theta_{I=0}$) 
and $J^\pi=3/2^-$ ($\Theta_{I=1}$) states
nearly degenerate. 
The widths of $\Theta^+_0$ and $\Theta^+_1$ are estimated 
to be $\Gamma < 7$ MeV and $\Gamma < 1$ MeV, respectively.
Two spin-zero diquarks are found in the $\Theta^+_0$, which confirms
Jaffe-Wilczek picture.
The origin of the novel level structure 
is the $5q$ dynamics of the confined system 
bounded by the connected flux-tubes.
We consider that the present results 
for the $J^\pi=\{1/2^+,3/2^+\}$$(\Theta^+_{I=0}$) states correspond 
to the experimental observation of $\Theta^+$,
while the $\Theta_{I=1}$ is not observed yet.
The existence of many narrow states,
$J^\pi=1/2^+$, $3/2^+$, and $3/2^-$, 
may give an light to further experimental observations.

Concerning other pentaquarks, we give a comment on $\Xi(ddss\bar{u})$.
The AMD calculations indicate that the diquark structure disappears in 
the $ddss\bar{u}$($1/2^+$) due to the $SU(3)$-symmetry breaking
in the color-magnetic interaction.
As a result, the estimated width of the $ddss\bar{u}$($1/2^+$) state 
is $\Gamma\approx 100$ MeV, which is much broader than  
$\Theta^+$($1/2^+$). Also the $3/2^-$ state is not so narrow because 
a $S$-wave decay channel $\Xi^*(1530)\pi$ is open. 

Finally, we would like to remind the readers that the 
absolute masses of the pentaquark in the present work are not predictions. 
We have an ambiguity of the zero-point energy of the string potential,
which depends on the flux-tube topology in each of meson, $3q$-baryon, 
pentaquark systems. 
We adjust that for the pentaquarks to reproduce the observed $\Theta^+$ mass. 
To confirm the zero-point energy,
experimental information for other pentaquark states 
are desired.

The authors would like to thank to 
T. Kunihiro, 
Y. Akaishi and 
H. En'yo 
for valuable discussions. 
This work was supported by Japan Society for the Promotion of 
Science and Grants-in-Aid for Scientific Research of the Japan
Ministry of Education, Science Sports, Culture, and Technology.

\def\Ref#1{[\ref{#1}]}
\def\Refs#1#2{[\ref{#1}\ref{#2}]}
\def\npb#1#2#3{{Nucl. Phys.\,}{\bf B{#1}}\,(#3)\,#2}
\def\npa#1#2#3{{Nucl. Phys.\,}{\bf A{#1}}\,(#3)\,#2}
\def\np#1#2#3{{Nucl. Phys.\,}{\bf{#1}}\,(#3)\,#2}
\def\plb#1#2#3{{Phys. Lett.\,}{\bf B{#1}}\,(#3)\,#2}
\def\prl#1#2#3{{Phys. Rev. Lett.\,}{\bf{#1}}\,(#3)\,#2}
\def\prd#1#2#3{{Phys. Rev.\,}{\bf D{#1}}\,(#3)\,#2}
\def\prc#1#2#3{{Phys. Rev.\,}{\bf C{#1}}\,(#3)\,#2}
\def\prb#1#2#3{{Phys. Rev.\,}{\bf B{#1}}\,(#3)\,#2}
\def\pr#1#2#3{{Phys. Rev.\,}{\bf{#1}}\,(#3)\,#2}
\def\ap#1#2#3{{Ann. Phys.\,}{\bf{#1}}\,(#3)\,#2}
\def\prep#1#2#3{{Phys. Reports\,}{\bf{#1}}\,(#3)\,#2}
\def\rmp#1#2#3{{Rev. Mod. Phys.\,}{\bf{#1}}\,(#3)\,#2}
\def\cmp#1#2#3{{Comm. Math. Phys.\,}{\bf{#1}}\,(#3)\,#2}
\def\ptp#1#2#3{{Prog. Theor. Phys.\,}{\bf{#1}}\,(#3)\,#2}
\def\ib#1#2#3{{\it ibid.\,}{\bf{#1}}\,(#3)\,#2}
\def\zsc#1#2#3{{Z. Phys. \,}{\bf C{#1}}\,(#3)\,#2}
\def\zsa#1#2#3{{Z. Phys. \,}{\bf A{#1}}\,(#3)\,#2}
\def\intj#1#2#3{{Int. J. Mod. Phys.\,}{\bf A{#1}}\,(#3)\,#2}
\def\sjnp#1#2#3{{Sov. J. Nucl. Phys.\,}{\bf #1}\,(#3)\,#2}
\def\pan#1#2#3{{Phys. Atom. Nucl.\,}{\bf #1}\,(#3)\,#2}
\def\app#1#2#3{{Acta. Phys. Pol.\,}{\bf #1}\,(#3)\,#2}
\def\jmp#1#2#3{{J. Math. Phys.\,}{\bf {#1}}\,(#3)\,#2}
\def\cp#1#2#3{{Coll. Phen.\,}{\bf {#1}}\,(#3)\,#2}
\def\epjc#1#2#3{{Eur. Phys. J.\,}{\bf C{#1}}\,(#3)\,#2}
\def\mpla#1#2#3{{Mod. Phys. Lett.\,}{\bf A{#1}}\,(#3)\,#2}
\def\etal{{\it et al.}}

\end{document}